\documentclass{aastex}
\usepackage{emulateapj5}

\newcommand{\ms}{\:{\rm ms}}
\newcommand{\km}{\:{\rm km}}
\newcommand{\ks}{\:{\rm ks}}
\newcommand{\ev}{\:{\rm eV}}
\newcommand{\kev}{\:{\rm keV}}
\newcommand{\y}{\:{\rm y}}
\newcommand{\es}{\:{\rm erg/s}}
\newcommand{\Gauss}{\:{\rm G}}
\newcommand{\nH}{\ensuremath{\rm n_H}}
\newcommand{\pc}{\:{\rm pc}}

\newcommand{\chisq}{\ensuremath{\chi^2}}
\newcommand{\redchisq}{\ensuremath{\chi^2_\nu}}
\newcommand{\Thot}{\ensuremath{\rm T_{hot}}}
\newcommand{\Tcold}{\ensuremath{\rm T_{cold}}}

\newcommand{\XMM}{{\it XMM}}
\newcommand{\CXO}{{\it CXO}}
\newcommand{\ROSAT}{{\it ROSAT}}
\newcommand{\HST}{{\it HST}}
\newcommand{\EUVE}{{\it EUVE}}

\newcommand{\rxj}{\ensuremath{\rm RX\:J1856-3754}}

\newcommand{\NS}{ {\rm NS} }
\newcommand{\Mstar}{\ensuremath{\rm M_\NS}}
\newcommand{\Rstar}{\ensuremath{\rm R_\NS}}
\newcommand{\Pstar}{\ensuremath{\rm P_\NS}}
\newcommand{\Rinf}{\ensuremath{\rm R_\infty}}
\newcommand{\Thrad}{\ensuremath{\Theta_{\rm rad}}}
\newcommand{\Msun}{ \ensuremath{\rm M_\odot} }

\newcommand{\ie}{{\it i.e.}}
\newcommand{\eg}{{\it eg.}}
\newcommand{\et}{{\it et~al.}}
\newcommand{\vs}{{\it vs.}}

\newcommand{\fig}[1]{Figure~\ref{#1}}
\newcommand{\sect}[1]{\S\ref{#1}}
\newcommand{\tbl}[1]{Table~\ref{#1}}

\newenvironment{inlinefigure}{
\smallskip
\def\@captype{figure}
\noindent\begin{minipage}{0.999\linewidth}\begin{center}}
{\end{center}\end{minipage}\smallskip}

\slugcomment{\apj \, in press.}
\shorttitle{\rxj: Evidence for a Stiff EOS}
\shortauthors{Braje \& Romani}

\begin{document}

\title{\rxj: Evidence for a Stiff EOS}

\author{Timothy M. Braje and Roger W. Romani}
\affil{Physics Department, Stanford University, Stanford, CA 94305}
\email{timb@astro.stanford.edu, rwr@astro.stanford.edu}

\begin{abstract}

We have examined the soft X-ray plus optical/UV spectrum of the nearby
isolated neutron star \rxj, comparing with detailed models of a
thermally emitting surface. Like previous investigators, we find the
spectrum is best fit by a two-temperature blackbody model. In
addition, our simulations constrain the allowed viewing geometry from
the observed pulse fraction upper limits. These simulations show that
\rxj\ is very likely to be a normal young pulsar, with the non-thermal
radio beam missing Earth's line of sight. The SED limits on the model
parameter space put a strong constraint on the star's $M/R$. At the
measured parallax distance, the allowed range for $\Mstar=1.5\Msun$ is
$\Rstar=13.7\pm0.6\km$. Under this interpretation, the EOS is
relatively stiff near nuclear density and the `Quark Star' EOS posited
in some previous studies is strongly excluded. The data also constrain
the surface $T$ distribution over the polar cap.
\end{abstract}


\section{Introduction}

\rxj, discovered by \citet{walt96}, is the nearest and brightest known
neutron star not showing emission dominated by non-thermal
magnetospheric processes. As such it offers a unique opportunity to
study the bare thermal surface emission. Measurements of the spectrum
can probe the neutron star mass (\Mstar) and radius (\Rstar),
constraining the high density equation of state (EOS). Since the
discovery, there have been several intensive observing campaigns
covering the optical-UV (\HST) and most recently the detailed soft
X-ray spectrum (\CXO). An initial $50\ks$ Low-Energy Transmission
Grating (LETG) spectrum showed a broad band spectrum remarkably
consistent with a simple blackbody \citep{bur01}, although hints of
spectral features were suggested \citep{vanKer02}. A deeper
observation using $450\ks$ of Director's Discretionary Time (DDT) was
made. This unique data set has been the subject of prompt study;
several authors show that lines in the spectrum are undetectable,
while pulse searches have placed stringent limits on the observed soft
X-ray pulse fraction \citep{ran02,dra02}. These data have been
variously interpreted, including the widely reported speculation
(based on the X-ray spectrum alone) that \rxj\ might be a bare quark
star \citep[esp.][]{dra02}.

Despite the very stringent constraints placed on the X-ray pulse
fraction, \citep[$<4.5\%$ at $99\%$ confidence, including a ${\dot P}$
search][]{ran02}, there is strong evidence that \rxj\ is a
rotation-powered pulsar. \citet{vanKer01b} discovered an H$\alpha$
nebula surrounding the neutron star, concluding that it could be best
interpreted as a bow-shock nebula powered by a relativistic wind of
$e^\pm$ generated by pulsar spindown. The bow shock geometry then
provides an estimate of the spindown power $\dot{E} = I \Omega {\dot
\Omega} = 8\times 10^{32}\es~d_{140}^3$ \citep{vanKer01b}. Adopting
magnetic dipole braking at constant $B$, this gives $\dot{E} = 10^{34}
(B_{12} \tau_6)^{-2}\es$ for a surface dipole field
$10^{12}B_{12}\Gauss$ and characteristic age $10^6\tau_6\y$,
suggesting $B_{12} \tau_6 \sim 3$.

A critical parameter in the discussion of this source is the distance,
which has been the subject of some controversy. Initial estimates from
\HST\ astrometry gave a parallax distance of $61\pc$ \citep{wal01}.
\citet{kap02}, however re-analyzed these data, deriving $d=143\pc$. A
fourth \HST\ observation appears to have resolved this discrepancy,
giving an overall measurement of $d=117\pm 12\pc$ \citep{wal02}; we
adopt this value.

\section{Spectral Fits\label{specfits}}

For some time now, it has been clear that the broad band spectrum of
\rxj\ from \HST, \ROSAT, and \EUVE\ data \citep[for a detailed
discussion, see][]{pons02} is inconsistent with a light element
($\sim$ Kramer's law opacity) H or He atmospheres. H models, for
example, overpredict the optical/X-ray flux ratio by a factor
$\sim100$ \citep{pav96}. Blackbody, heavy element or composite models
gave acceptable fits. To produce such $\sim$ Planckian spectra, one
must have nearly isothermal conditions at optical depth $\tau_\nu
\approx 1$ across the observed band. One possibility is that the
surface is in a solid or liquid state, precluding large temperature
gradients through the photosphere. Theoretical studies to date, while
limited, suggest that this is not the case unless H is present and
$B>10^{13}\Gauss$ \citep{lai97}. In an atmosphere, the easiest way to form
the spectrum in a small depth range is to invoke a rich line spectrum;
thus, at first sight, blackbody-like spectra should require a high
spectral density of opacity features (lines and edges). This led to
the expectation that the blackbody-like spectrum of \rxj\ would show
many spectral features when examined with good S/N LETG spectra.
Unfortunately, the initial $50\ks$ exposure already showed that the
line features were substantially weaker than expected in a simple
low-$B$ single temperature atmosphere model dominated by heavy
elements. The new DDT exposure only enhances this conclusion, placing
very strong limits \citep[typical equivalent width
$\la0.02$\AA][]{dra02} on spectral features.

We will be concerned with the quality of statistical fits to various
atmosphere models. It is important to note here that blackbody
spectral fits to the \CXO\ \rxj\ LETG data are, contrary to early
reports, \emph{not} statistically perfect fits to a simple Planck spectrum.
This conclusion was drawn from fits to basic CIAO extractions, which
appreciably underbin the spectrum. Instead we find that at a more
modest binning (equally spaced $\sim0.7$\AA bins) we obtain
$\chi^2/$DOF=1.6. As the bin size is increased, the $\chi^2/$DOF grows
to $\sim 4.8$, until the number of degrees of freedom becomes small.
This is a clear signature of spectral departures on resolved energy
scales, and with appropriate binning one indeed finds systematic,
grouped residuals to the Planck function fit at the $\sim 10$\% level.
We believe that these represent the limit of accuracy in calibration
of the response matrix, as the broad band spectral shape is an
excellent fit to the Planck function. \citet{dra02} have reached
similar conclusions. Recognizing that very subtle departures from a
pure blackbody may be present in these data, we adopt the conservative
assumption that these departures are fully accounted for by response
matrix systematics.

To accommodate an extended atmosphere, one must suppress the spectral
features. One possibility is that external heat sources (such as
precipitating magnetosphereic $e^\pm$) drive the atmosphere towards
isothermality. Sample atmospheres showing this effect have been
computed in \citet{gan02}. A second possibility is that the line
energies for a given species vary strongly across the neutron star
surface. For normal pulsar fields, $B \sim 10^{12}\Gauss$ or higher,
the strong dependence of the transition energies on the local $B$
\citep[\eg][]{rrm97,pav95}, coupled with substantial $\ge 2\times$
variation of $B$ across the surface, even for the simplest dipole
models, ensures that such `magnetic smearing' will strongly suppress
the phase-averaged line width (Romani, in prep.). We discuss briefly
here a third possibility, that the lines experience variable shifts as
the pulsar rotates due to Doppler and other dynamic effects. Again,
the phase-averaged spectrum observed for \rxj\ would be expected to
show broadened and blended lines, driving the spectrum towards a
Planck curve.

If, as is required for a normal neutron star, the soft X-ray emission
of \rxj\ is dominated by hot polar caps, the rotation of these past
the line of sight produces phase dependent Doppler shifts
\citep{bra00}. These are only significant for $v_{surf} \sim 2 \pi
\Rstar/\Pstar \longrightarrow c$, \ie\ $\Pstar \la$ a few ms. Such
small $\Pstar$ are not excluded, since the HRC-S wiring error limits
arrival time accuracy and precludes sensitive searches for $\Pstar \la
10\ms$. Moreover, the ${\dot E}$ from the bow shock standoff gives
$\Pstar =4.6 (B/10^8\Gauss)^{1/2}\ms$, so a low field star, having a
non-magnetic atmosphere would have a $\sim\ms$ spin period.

For a concrete example, we assume $\Mstar = 1.4\Msun$, 
$\Rstar=10\km$, and $\Pstar=1.5\ms$ (allowed, as the existence of
$PSR~1937+21$ shows).  We have tried both solar
abundance and iron model atmospheres, tested a range of magnetic inclinations
$\alpha$, and computed phase averaged spectra 
using an extension of the Monte Carlo simulation code described in
\citet{bra00}.  We then tested these models, fitting the most recent \CXO\ 
data, allowing the temperature, observer viewing angle $\zeta$, and interstellar 
absorption \nH\ to vary.  The large L and M shell edges in the iron models
ensure that these are always poor fits. The solar abundance atmospheres
have a much richer line structure which is more easily blurred into a 
pseudo-continuum.  In \fig{solfit}, we display the solar abundance 
millisecond pulsar model, overlaid on the \CXO\ data. For comparison the best fit
non-rotating model is shown in the upper panel.  Doppler boosting 
produces a qualitatively acceptable fit below $\sim0.5\kev$, but the simple 
blackbody remains statistically an appreciably better model ($\chi^2$/DOF=1.6
\vs\ 3.7 for the binning chosen). We must conclude that a simple blackbody
fit remains the best available, although not yet physically explained. 
Planck emission from a physical neutron star can be compared with the data to
obtain significant constraints on the neutron star parameters; we pursue
this in the remainder of the paper.

\begin{inlinefigure}
\scalebox{0.75}{\rotatebox{270}{\plotone{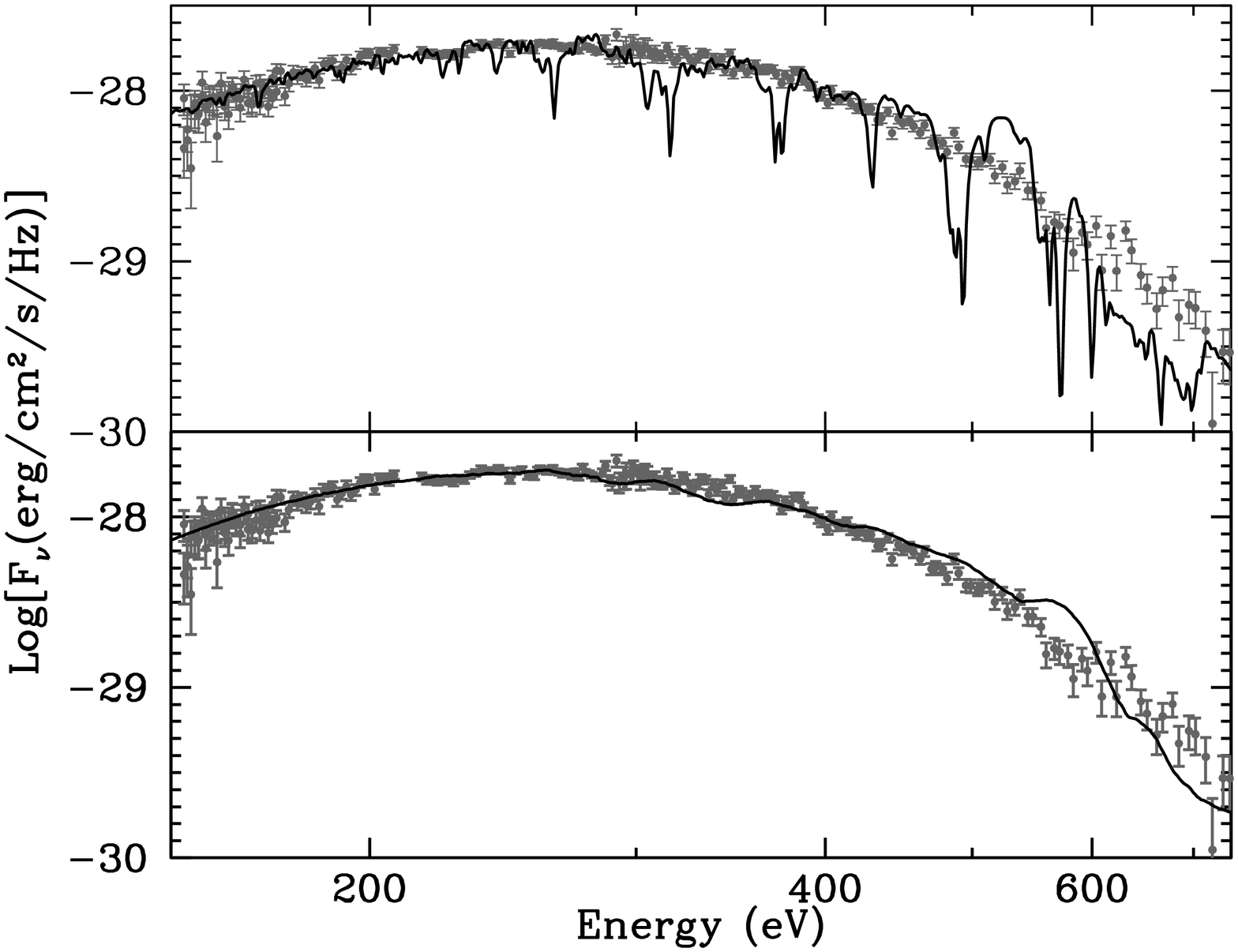}}}
\figcaption{\emph{Top:} Best fit solar abundance model with no Doppler shifts.
\emph{Bottom:} Best fit solar abundance model with Doppler shifts for a $\Pstar=1.5\ms$, 
$\Rstar=10\km$ pulsar. \label{solfit}}
\end{inlinefigure}

\section{Two Temperature Blackbody Fits}

A blackbody fit to the \CXO\ data alone results in
a temperature of $T_\infty \sim 61\ev$ with very small
statistical uncertainty \citep[][and our own analysis]{dra02}. 
We find, as also reported by \citet{dra02}, that
systematics (likely in the the effective area, as noted above) provide
the dominant error. \citet{dra02} quote $61.2\pm1.0\ev$. Taken at face
value this  $T_\infty$ with the parallax distance gives a
radius as measured at infinity of $\Rinf=(1+z)\Rstar=3.8-8.2\km$. 
If interpreted as the full star radius, this demands exotic equations of state 
(\ie\ quark stars). Of course, this is only a lower limit to the stellar radius.
The optical/UV data points, which closely follow a Rayleigh-Jeans spectrum
\citep[\eg][]{pons02, vanKer01a}, are most easily interpreted as
a second cooler Planck spectrum representing flux from the full surface.
This has been previously recognized, but \citet{dra02} argue against this
interpretation, citing the absence of the X-ray pulse expected from
such a hot polar cap/cool surface combination. We have addressed this
concern quantitatively, computing detailed light curves and spectra.

\subsection{Analytic Two-Temperature Models}

A simple analytic two-temperature blackbody fit delineates the basic model
parameters.  For a range of effective (cold surface) radii, we fit
$\Thot$, $\Tcold$, the hot area, and the absorption column density.
In Figure 2, we plot the minimum \chisq\ as a function of cold radius. 
The optical-UV data points fix $\Rstar^2\Tcold$.
The fit becomes poor when $\Tcold$ starts to allow significant Wien peak 
contribution to the \CXO\ X-ray band; this sets the minimum stellar radius.
Formally, there is a maximum acceptable radius beyond which low $\Tcold$
predicts Wien peak curvature inconsistent with the $\sim$Rayleigh-Jeans
UV data points.

\begin{inlinefigure}
\scalebox{.75}{\rotatebox{270}{\plotone{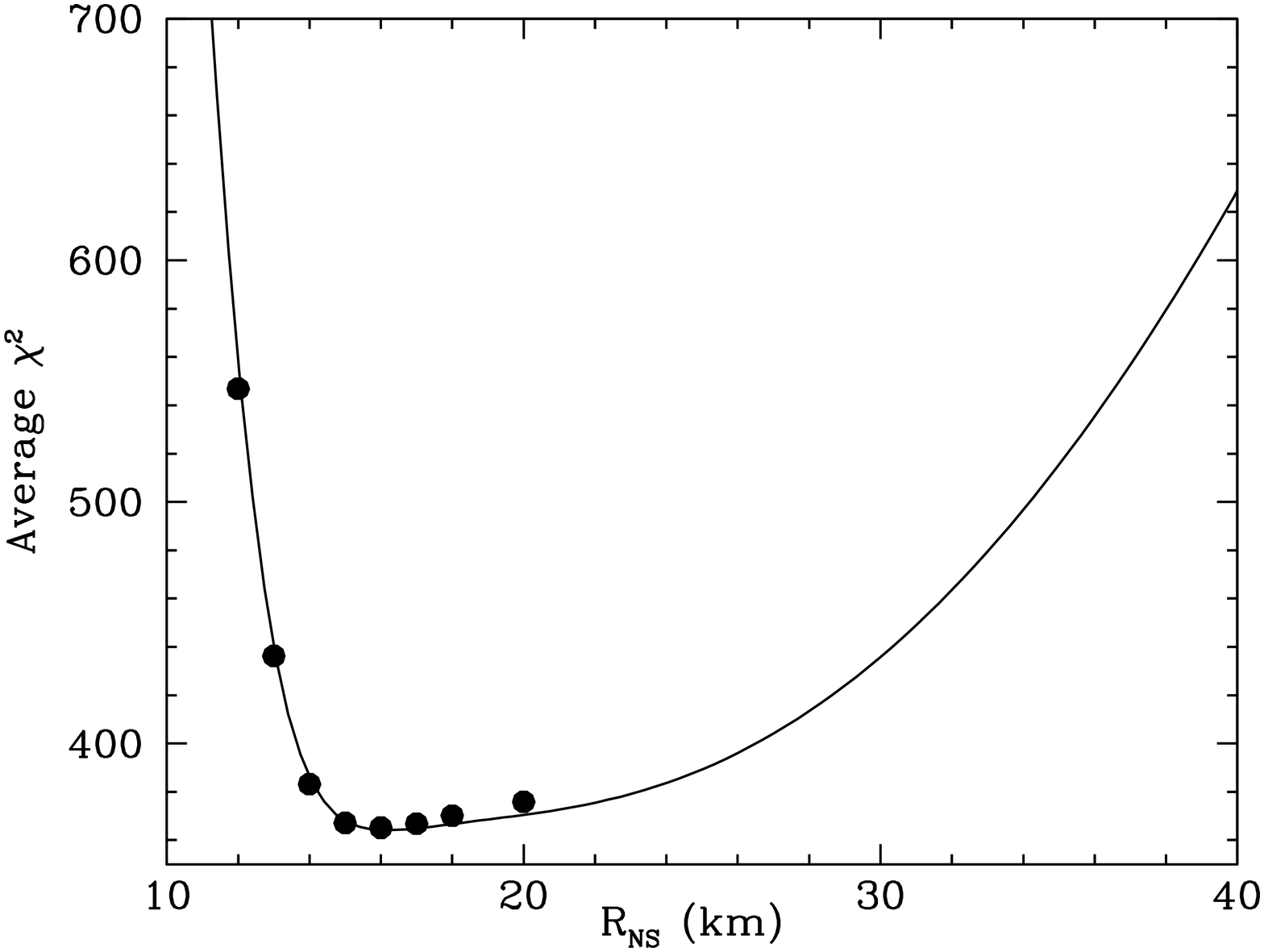}}}
\figcaption{Solid line: \chisq\ (257 DOF) as a function of cold sphere 
radius.  Points: \chisq\ values from the polar-cap model fits.}
\label{chi}
\end{inlinefigure}

\subsection{More Realistic Two-Temperature Models}

In addition to the phase averaged spectrum, we have a limit
on the \CXO-band pulse fraction. A more realistic model is required to
address the detailed spectrum and pulsations.  We adopt a two-temperature model
with two opposing hot spots (polar caps) at $\Thot$ and the remainder
of the surface at $\Tcold$. The caps' orientation ($\alpha$ and $\zeta$)
are free parameters.  We radiate from these surfaces,
tracing the photons to infinity to form phase resolved spectra
and light curves. For details of these Monte-Carlo sums see \citet{bra00}.

The analytic model results allow some useful simplifications.
Since $\Thot$ is virtually constant over the full acceptable $\Rstar$
range, we fix this value in exploring the rest of the parameter space.
Further, the X-ray flux amplitude allows an initial estimate of 
the cap half angle $\Delta$ (which depends on $\alpha$ and $\zeta$) that 
ensures that the pulse formation is accurate and results 
in quick convergence to the true minimum.

The fit parameters of interest are $\Rstar$, $\Delta$, 
$\Tcold$, $\zeta$, $\alpha$, and $\nH$. 
To explore the sensitivity to several parameters, we have computed a model 
grid.  We calculate all models for $\alpha=5\arcdeg$ to $\alpha=90\arcdeg$ in 
five degree steps; $\zeta=0\arcdeg$ to $\zeta=90\arcdeg$ in five degree steps; 
and $\Rstar=12\km$ to $\Rstar=20\km$ in one kilometer steps. While not
strictly a fit parameter, we also vary the stellar mass $\Mstar$.
The quality of the spectral fit turns out to be
quite insensitive to the choice of $\alpha$ and $\zeta$.
In Figure 3, we display a typical spectrum for a model with 
$\Mstar=1.4\Msun$, $\Rstar\ga14\km$.
As the radius becomes smaller, the optical-UV flux is under-predicted and
the fit becomes unacceptable.

\begin{inlinefigure}
\scalebox{.75}{\rotatebox{270}{\plotone{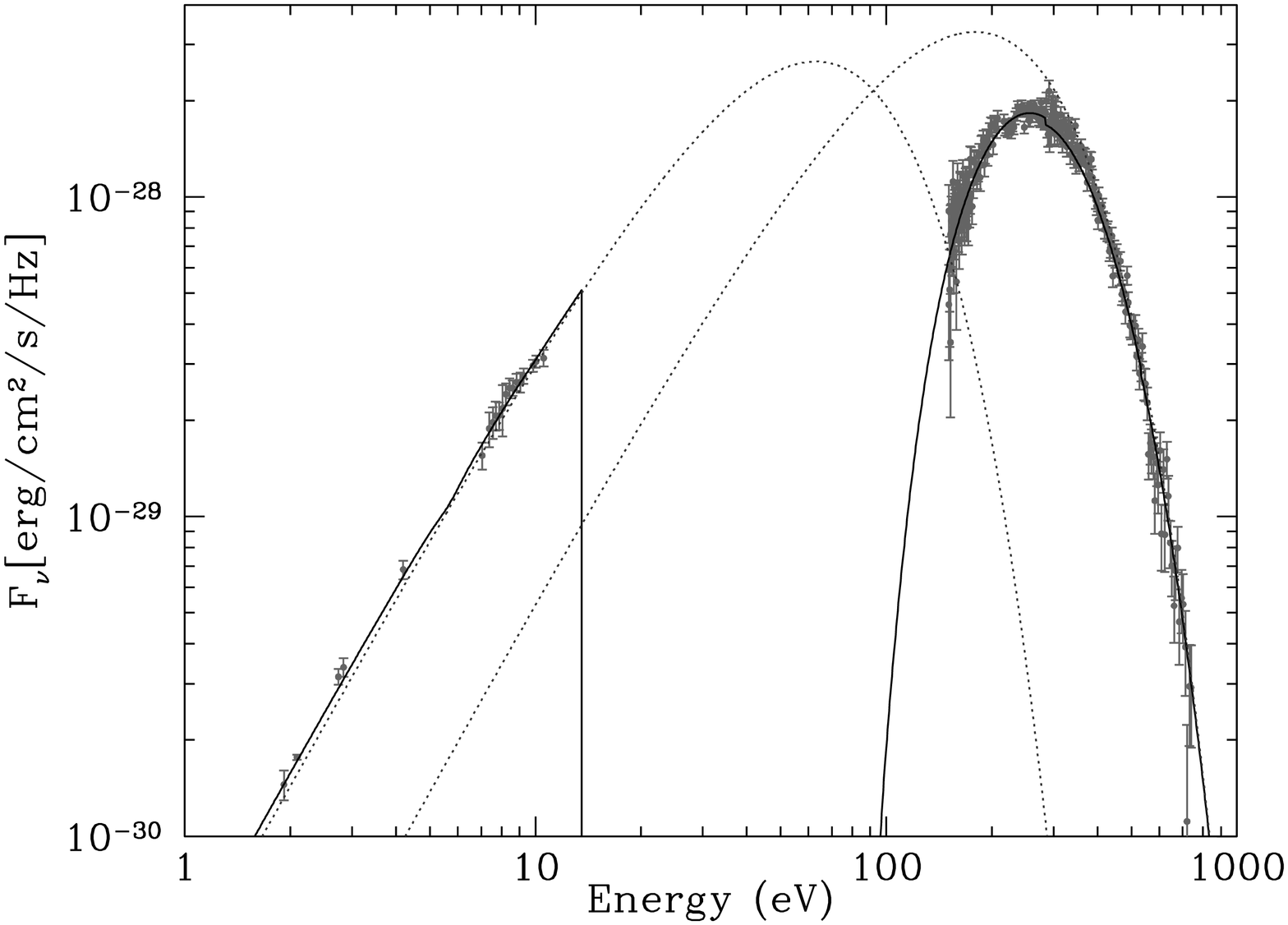}}}
\figcaption{Broad band spectral fit to \rxj. Optical/UV data points are drawn from
\citet{vanKer01a} and \citet{pons02}.  The dotted lines show the unabsorbed hot and cold
blackbody components.}
\label{specfit}
\end{inlinefigure}

For each $\alpha$ and $\zeta$ we compute \chisq\ as a function of $\Rstar$.  
The average over angles is shown by the points in Figure 2.
Note that the minimum value is not at $\redchisq=1$, a consequence of the 
aforementioned systematic errors. This means that the errors
are not Poisson distributed according to the bin counts. To establish
confidence levels (CL), we must Monte Carlo according to the observed
error distribution. We have computed the \chisq\ distribution
about the best fit model for each ($\alpha$, $\zeta$) combination, obtaining
a histogram of \chisq\ values. These were almost completely insensitive
to the angles, so we combined all the \chisq\ distributions to obtain
confidence levels to the \chisq\ increases associated with variations
in \Rstar. Examinations of the differences between the different
angles and between independent Monte Carlo runs show that the estimates
of 90\% and 99\% confidence level limits on the radius are uncertain by no more than
$\pm0.5\km$ from systematic and computational errors in this procedure.

\subsection{Cap Shape Constraints}

One might question whether a simple, circular uniform $\Thot$
polar cap is merely an adequate approximation to the data.  We have fit
some alternative surface temperature distributions $T(\eta)$, where
$\eta$ is the magnetic co-latitude, comparing with the best fit circular
cap which had $k\Thot=62.8\ev$, $\Delta=21\arcdeg$. The best fit
Gaussian $T(\eta)$ had a peak temperature of $k\Thot=74.8\ev$ and 
width $\sigma=19\arcdeg$, but showed an increase of $\chisq \approx 80$
over the simple cap model, sufficient to exclude at the $\sim 99$\% CL.
Adding a simple linear $\Thot$ to $\Tcold$ ramp to a uniform cap
makes no discernible difference until the ramp width is twice that of the cap.
At this point the best fit model has a cap with $k\Thot = 69\ev$ and 
$\Delta \approx 12\arcdeg$, but the model is excluded at the $\sim 90$\% CL.
Finally, we fit both a simple $\Thot \propto \cos(\eta)$ model
and the surface $T(\eta)$ distribution of \citet{gre83}
which is motivated by magnetic anisotropy in the thermal conductivity. Both
were excluded by the \CXO\ data at very high confidence. All fits were
at $1.4\Msun$, best fit stellar radii and ($\alpha$, $\zeta$) chosen so that
the pulse fraction is $\la 5\%$ for the default cap model.
 
Evidently, the \CXO\ data require a quite uniform distribution of 
the high temperature excess, and suggests that it is induced by exterior 
heating rather than interior conductivity. This cap size is substantially
larger than the $\Delta \sim 3\arcdeg(100\ms/\Pstar)^{1/2}$ expected for a dipole
surface cap, unless the period is very small. Higher magnetic multipoles 
would generally have even smaller open zone caps. The large, uniformly
heated area is puzzling in the context of pulsar surface acceleration
models, but might be most easily accommodated in the more modern
pictures of a GR-induced potential that is relatively
uniform across the polar cap and forms a pair formation
front at relatively high altitudes \citep{hm98}. Perhaps a more
plausible interpretation invokes a high-altitude acceleration zone,
with the inward-directed $\gamma$-rays pair converting in the closed zone
above the polar cap \citep{wrhz98}, which should give a $\sim$ large
zone of uniform surface heating.

\section{Pulse Fraction Constraints}

We have seen that the spectral fits are quite insensitive to 
the cap viewing angles. They place a firm minimum on the allowed
\Rstar\ but allow radii that are implausibly large.
However, the observed pulse fraction depends strongly
on viewing geometry and, through gravitational focusing, the
value of $\Mstar/\Rstar$. In general, we expect a strong X-ray pulse when
the magnetic axis passes close to Earth's line-of-sight (small
$|\alpha-\zeta|$). Of course, for an aligned rotator $\alpha\approx0$
the pulse can be very small and for $\alpha\approx\pi/2$ there is
an appreciable region where the pulse is weak as viewed from Earth.
We have examined the \CXO-band pulse profile of our model grid to find
the allowed region of $(\alpha,\zeta)$ parameter space in which
the model pulse fraction is weaker than that observed ($\la4.5\%$);
this is shown in the inset of \fig{skycoverage}.

This exercise is repeated for each $\Mstar,\Rstar$; the allowed
phase space is larger for small $\Rstar/\Mstar$ as gravitational bending dilutes
the observed pulse. By demanding a certain minimum probability that a
pulse fraction as low as that observed is seen at Earth, we obtain
a {\it maximum} acceptable radius for the neutron star. One important
additional constraint can be invoked. The non-detection of this neutron 
star in the radio band \citep{bra99} puts a very strong bound on pulsar
emission directed towards Earth. Assuming that this is a normal radio pulsar
(consistent with ${\dot E}$ inferred from the bow shock),
we must conclude {\it a priori} (independent of the
X-ray data) that our line of sight lies outside of the pulsar radio beam
and that $|\alpha-\zeta|$ is not small. The region excluded depends on
the size of the radio beam (\Thrad), which in turn depends on the spin period.
A typical estimate is \citep[\eg][]{ran93}
\begin{equation}
  \Thrad = 5.8\arcdeg (\Pstar/1s)^{-1/2}
\end{equation}
at the radio frequency $\nu=1$~GHz.  With $\Pstar=0.3~{\rm s}$, this gives 
$\Thrad \sim 10.6\arcdeg$.  Radio limits on \rxj\ are strong at even 
lower $\nu$ where radius-to-frequency mapping produces larger $\Thrad$,
and for this nearby object, the radio luminosity constraints are so severe
that we are unlikely to intersect even the faint fringe of the radio
beam. Both effects argue for $\Thrad$ larger than that above.
Accordingly, in the \fig{skycoverage} inset we show by diagonal lines the
regions near $\alpha=\zeta$ excluded by the lack of radio detection for
$1\times$ and $2\times$ this fiducial beam size. Shorter periods
allow an {\it a priori} exclusion of even more phase space. 
The point is that from the lack of radio detection we should have {\it expected}
a low X-ray pulse fraction. The probability for obtaining pulses as weak as 
observed is then set by the fraction of the remaining allowed solid angle.
These fractions are plotted in \fig{skycoverage} with and without the radio 
prior.  The allowed radius range depends on mass and in \tbl{tbl-pf}, we 
give the 90\% and 99\% CL upper bound on the neutron star radius
for these priors and several neutron star masses.

\begin{inlinefigure}
\scalebox{0.9}{\rotatebox{0}{\plotone{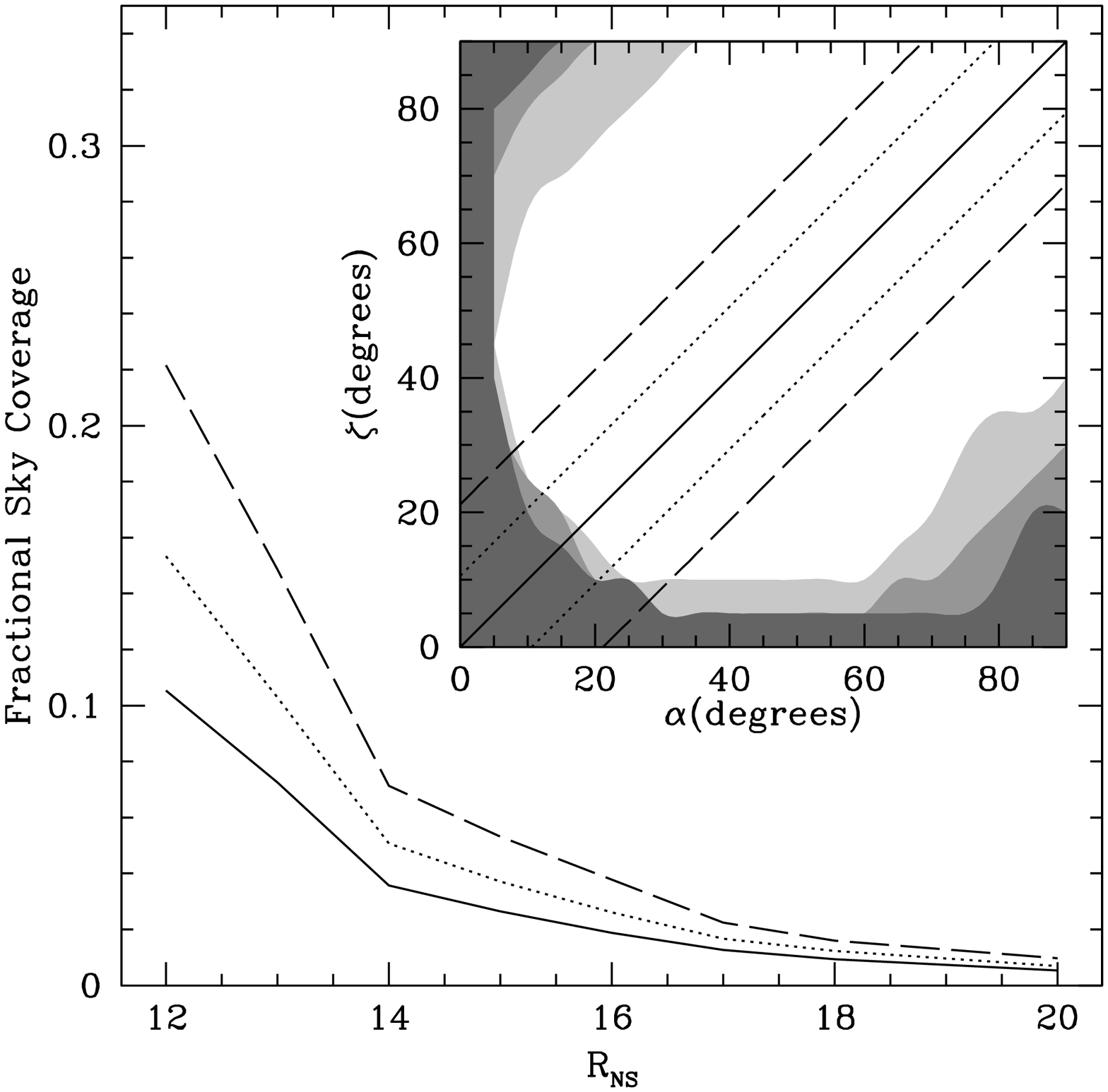}}}
\figcaption{Fraction of the sky allowed by pulse fraction constraints
as a function of stellar radius. The solid line assumes no priors; the
dotted line for a prior $1 \times \Thrad = 10.6\arcdeg$; and the
dashed line for $2 \times \Thrad$. The inset shows the allowed
($\alpha$,$\zeta$) parameter space shaded in light, medium, and dark
gray for $12, 14, 16\km$, respectively at $\Mstar=1.4\Msun$.  The
lines in the inset depict the parameter space excluded by radio prior.}
\label{skycoverage}
\end{inlinefigure}

\section{Equation of State Constraint}

We see that the X-ray/optical data, using the spectral and pulse 
fraction arguments, give a range of allowed stellar radii for each mass.
Strictly speaking, the minimum and maximum radius CL have somewhat different
interpretations, but it is interesting to place these bounds in 
the $\Mstar-\Rstar$ plane to compare with the predictions of various EOS.

In \fig{eosconstraint}, we show the combined constraints,
assuming $\Thrad = 21.2\arcdeg$. The spectral
radius lower bounds at 90\% and 99\% CL approximate curves of constant 
$\Mstar/\Rstar$. The pulse fraction upper bounds (90\%, 95\%) rapidly 
drive one to small radii at low masses.  For $\Mstar\la1.3\Msun$ no simultaneous
solutions are allowed consistent with the 90\% bounds. At $\Mstar=1.5\Msun$
the allowed range from the fit is quite small, $\Rstar=13.7\pm0.6\km$.
The additional uncertainty in the distance actually dominates the errors (arrowed bar).

For comparison, several EOS curves \citep[after][]{lat01} are plotted.
We see that large radius (stiff at nuclear density) EOS are preferred.
Formally, the relativistic field theoretical model by \citet{mul96}
and the model GS2 by \citet{gle99} are the only modern models allowed
(the original PS model of \citet{pan75} is also allowed). We note that
the GS models are very sensitive to the K meson potential; for example,
GS1 is strongly excluded. Interestingly, no potential or variational
method computations agree with the formal overlap. If one includes the
distance uncertainty, a few more intermediate radius models are not
excluded at the 90\% CL. However, even including the distance
uncertainties, all quark star models are excluded at the $\sim95$\%
level for $\Mstar\la1.5\Msun$ and are only barely consistent at the
highest allowed masses. Improved \HST\ parallax measurements could
boost this exclusion to the 3$\sigma$ level.

\begin{inlinefigure}
\scalebox{0.9}{\rotatebox{0}{\plotone{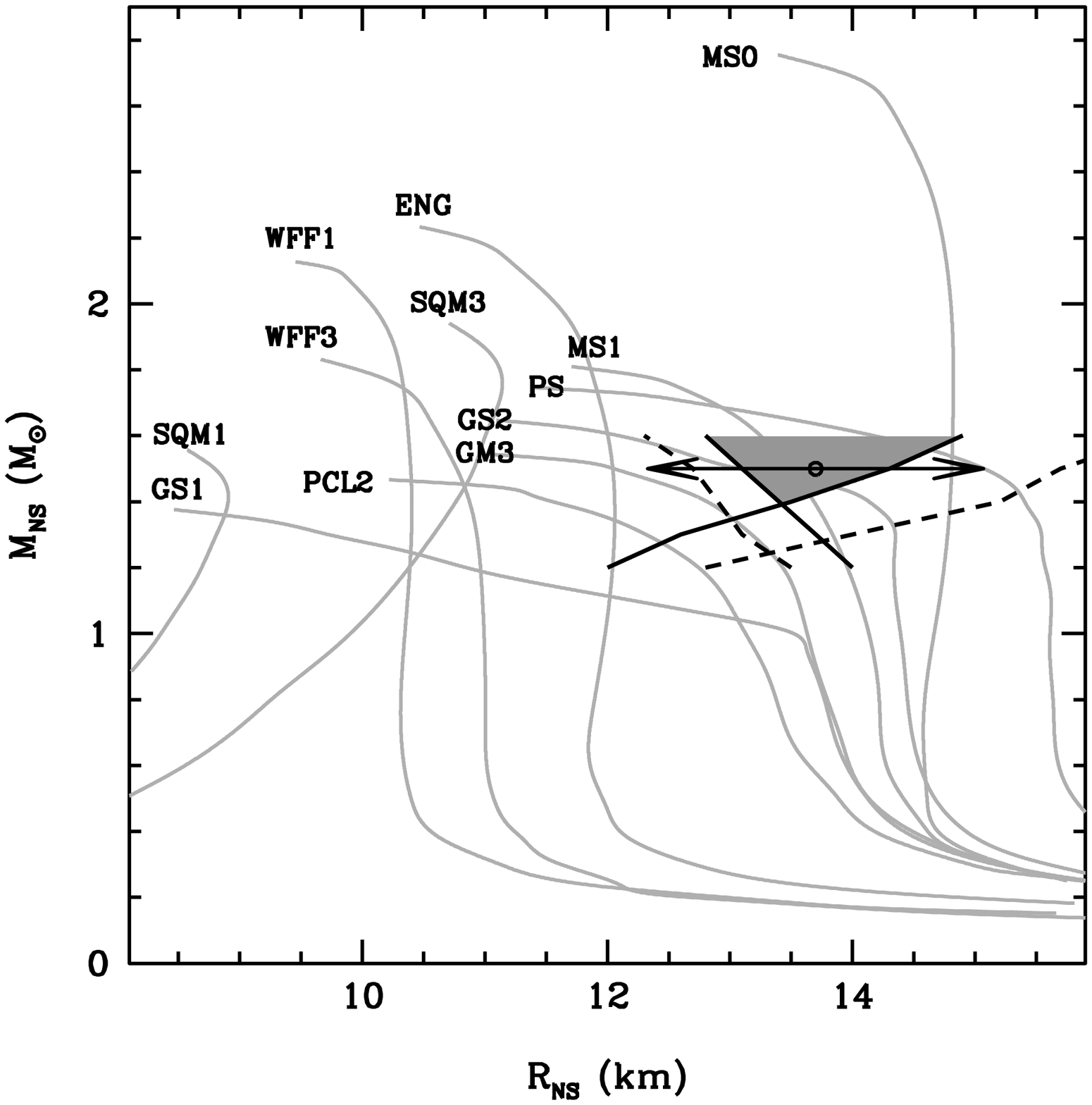}}}
\figcaption{Radius constraints for different possible \rxj\ masses.  The
triangular gray shaded region represents the formal 90\% CL
overlap from the pulse fraction and spectral constraints.
The two-sided arrow represents the systematic range induced by
the distance uncertainties. Equation of state curves and labels are drawn from
\citet{lat01}.  See this paper for EOS labels and references.}
\label{eosconstraint}
\end{inlinefigure}

The model GS2 differs from the MS models in that it includes a kaon
condensate in the core. One principal effect of exotic interior
condensates is to enhance neutrino cooling after $\sim100\y$.
This can also be achieved without exotica for a
proton fraction $\gtrsim 11\%$ through the direct URCA process
\citep{lat91}.  Our inferred $\Tcold$,
interpreted as the signature of the cooling from the initial heat of
formation, corresponds to $L \approx 3.2 \times 10^{31}\es$ for our
best fit radius. Such luminosities are reached in $\sim 5\times
10^5\y$ \citep[the preferred age of][]{wal02} when enhanced neutrino cooling can
occur, but are achieved after $\sim 1.3 \times 10^6\y$ for stars with
low density (stiff) cores \citet{tsu02}. Thus, depending on the actual
stellar age, this cool surface may be seen as weak evidence for an exotic
composition and/or significant softening of the EOS at very high
densities.

The prospects for further tests of the ideas in this paper hinge on
the detection of a pulse from \rxj\ along with measurement of the
pulse fraction and the period derivative. Given that significant
allowed pulse fraction parameter space lies just slightly below the
\CXO\ LETG detection threshold, the prospects for a pulse measurement
with the recently completed $58\ks$ \XMM\ observation are quite good.
Even if only thermal, the phase resolved spectrum should provide
important constraints on the cap temperature, size and orientation. In
particular our thermal model predicts that the pulse fraction should
increase by 70\% from $0.15\kev$ to $0.5\kev$, where the \XMM\ data should
still deliver $\sim 0.8$~PN camera cps, allowing detection of pulse
fractions well under 1\%.

\acknowledgments

We are grateful to Boris G\"ansicke for collaboration on the atmosphere
models used in \sect{specfits}; and to Herman Marshall for sharing an independent
LETG response matrix.  This work was supported in part by NASA grant SP2-2002X.

%\clearpage
\begin{deluxetable}{cccc}
\tabletypesize{\normalsize}
\tablecaption{Pulse Fraction $90\%/99\%$ Confidence Limits\label{tbl-pf}}
\tablehead{
\colhead{\Mstar} & \colhead{No Prior} & \colhead{$\Thrad = 10.6\arcdeg$} 
& \colhead{$\Thrad = 21.2\arcdeg$} \\
(\Msun) & ($\km$) & ($\km$) & ($\km$)
}
\startdata
1.2 & 10.3/14.6 & 11.3/15.6 & 12.0/16.3 \\
1.3 & 10.5/16.7 & 11.7/17.6 & 12.6/19.2 \\
1.4 & 12.2/17.8 & 13.1/18.9 & 13.6/19.9 \\
1.5 & 13.4/19.4 & 13.8/21.0 & 14.3/21.3 \\
1.6 & 14.0/20.0 & 14.6/21.4 & 14.9/21.2 \\
\enddata
\end{deluxetable}


\begin{thebibliography}{}
\bibitem[Braje, Romani, \& Rauch(2000)Braje, Romani, \& Rauch]{bra00} Braje, T.M.,
	Romani, R.W., \& Rauch, K.P.  2000, \apj, 531, 447
\bibitem[Brazier \& Johnston(1999)]{bra99} Brazier, K. T. S. \& Johnston, S.
        1999, \mnras, 305, 671
\bibitem[Burwitz \et(2001)]{bur01} Burwitz, V., Zavlin, V. E., Neuh\"{a}user, R., 
        Predehl, P., Tr\"{u}mper, \& Brinkman, A. C.  2001, \aap, 379, L35
\bibitem[Drake \et(2002)]{dra02}  Drake, J. J., Marshall, H. L., 
        Dreizler, S., Freeman, P. E., Fruscione, A., Juda, M., Kashyap, V., 
	Nicastro, F., Pease, D. O., Wargelin, B. J., \& Werner, K.   2002
        \apj, 572, 996
\bibitem[G\"{a}nsicke, Braje, \& Romani(2002)]{gan02} G\"{a}nsicke, B.T.
        Braje, T.M. \& Romani, R.W.  2002, \aap, 386, 1001
\bibitem[Glendenning \& Schaffner-Bielich(1999)]{gle99} Glendenning, N. K. \&
        Schaffner-Bielich, J.   1999, Phys. Rev. C, 60, 025803
\bibitem[Greenstein \& Hartke(1983)]{gre83} Greenstein, G. \& Hartke, G. J.  1983, \apj, 
        271, 283
\bibitem[Harding \& Muslimov(1998)]{hm98} Harding, A. K. \& Muslimov, A. G. 1998,
	\apj, 508, 328
\bibitem[Kaplan, van Kerkwijk, \& Anderson(2002)]{kap02} Kaplan, D. L., van Kerkwijk, M. H.,
        \& Anderson, J.  2002, \apj, 571, 447
\bibitem[Lai \& Salpeter(1997)]{lai97} Lai, D. \& Salpeter, E. E.  1997, \apj, 491, 270
\bibitem[Lattimer \et(1991)]{lat91} Lattimer, J. M., Pethick, C. J.,
  Prakash, M., \& Haensel, P.   1991, \prl, 66, 2701
\bibitem[Lattimer \& Prakash(2001)]{lat01} Lattimer, J. M. \& Prakash, M.
        2001, \apj, 550, 426
\bibitem[M\"uller \& Serot(1996)]{mul96} M\"uller, H. \& Serot, B. D.  1996, 
        Nucl. Phys. A, 606, 508
\bibitem[Pandharipande \& Smith(1975)]{pan75} Pandharipande, V. R. \& Smith, R. A.
        1975, Nucl. Phys. A, 237, 507
\bibitem[Pavlov \et(1995)]{pav95} Pavlov, G.G., Shibanov, Y.A., Zavlin, V.E., 
        \& Meyer, R.D.  1995, in The Lives of Neutron Stars, ed. M. A. Alpar, 
	U. Kiziloglu, \& J. van Paradijs (Dordrecht: Kluwer), 71-90
\bibitem[Pavlov \et(1996)]{pav96} Pavlov, G. G., Zavlin, V. E., Tr\"umper, J.,
        \& Neuh\"auser, R.   1996, \apjl, 472, L33
\bibitem[Pons \et(2002)]{pons02} Pons, J. A., Walter, F. M., Lattimer, J. M., 
        Prakash, M., Neuh\"{a}user, R., \& An, P.  2002, \apj, 564, 981
\bibitem[Rajagopal, Romani, \& Miller(1997)]{rrm97} Rajagopal, M.,
	Romani, R. W., \& Miller, M. C.  1997, \apj, 479, 347 
\bibitem[Rankin(1993)]{ran93} Rankin, J. M.  1993, \apj, 405, 285
\bibitem[Ransom, Gaensler, \& Slane(2002)]{ran02} Ransom, S. M., Gaensler, B. M., 
        \& Slane, P. O.  2002, \apjl, 570, L75
\bibitem[Tsuruta \et(2002)]{tsu02} Tsuruta, S., Teter, M.A., Takatsuka, T.,
	Tatsumi, T. \& Tamagaki, R. 2002, \apj, 571, L143
\bibitem[van Kerkwijk \& Kulkarni(2001a)]{vanKer01a} van Kerkwijk, M. H. \&
        Kulkarni, S. R.  2001, \aap, 378, 986
\bibitem[van Kerkwijk \& Kulkarni(2001b)]{vanKer01b} van Kerkwijk, M. H. \&
        Kulkarni, S. R.  2001, \aap, 380, 221
\bibitem[van Kerkwijk(2002)]{vanKer02} van Kerkwijk, M. H. 2002, 
        Proc. Jan van Paradijs Memorial Symposium, ed. Van den Heuvel, E. P. J., 
	Kaer, L., Rol, E., ASP, San Francisco
\bibitem[Walter, Wolk, \& Neuh\"auser(1996)]{walt96} Walter, F. M., Wolk, S. J., \&
        Neuh\"auser, R.   1996, \nat, 379, 233
\bibitem[Walter(2001)]{wal01} Walter, F. M. 2001, \apj, 549, 433
\bibitem[Walter \& Lattimer(2002)]{wal02}  Walter, F. M. \& Lattimer, J. 
        2002, astro-ph/0204199
\bibitem[Wang \et(1998)]{wrhz98} Wang, F.Y-H., Ruderman, M., Halpern,
	J. P. \& Zhu, T.  1998, \apj, 498, 373
\end{thebibliography}
\end{document}